\documentstyle[12pt,epsf]{article}

\def \dfrac #1#2 {\displaystyle\frac{#1}{#2}}
\def\be{\begin{eqnarray}}
\def\ee{\end{eqnarray}}
\def\bq{\begin{equation}}
\def\eq{\end{equation}}
\def\ben{\begin{enumerate}}\def\een{\end{enumerate}}

\def\roughly#1{\mathrel{\raise.3ex\hbox{$#1$\kern-.75em
\lower1ex\hbox{$\sim$}}}}

\newcommand{\lrpartial}{\raisebox{0.1em}{$
\stackrel{\raisebox{-0.03em}{$\scriptstyle\leftrightarrow$}}{\partial}$}{}}
\newcommand{\lpartial}{\raisebox{0.1em}{$
\stackrel{\raisebox{-0.03em}{$\scriptstyle\leftarrow$}}{\partial}$}{}}
\newcommand{\rpartial}{\raisebox{0.1em}{$
\stackrel{\raisebox{-0.03em}{$\scriptstyle\rightarrow$}}{\partial}$}{}}

\begin{document}
\def\bra{\langle }
\def\ket{\rangle }
\begin{titlepage}

\vspace{.3cm} \hfill {\large FTUV-04-0831 ; IFIC-04-46}
\vspace{.2cm}
\begin{center}
\ \\
{\LARGE \bf Relativity and constituent quark structure
\vspace{.3cm}
\\
in model calculations of
parton distributions}
\ \\
\ \\
\vspace{1.0cm}
{
Santiago Noguera$^{(a)}$, Sergio Scopetta$^{(a,b)}$ and Vicente Vento$^{(a)}$}
\vskip 0.5cm
{\it (a) Departament de Fisica Te\`orica,
Universitat de Val\`encia, 46100 Burjassot (Val\`encia), Spain
\\
{\it and Institut de F\'{\i}sica Corpuscular,
Consejo Superior de Investigaciones Cient\'{\i}ficas}}
\\
{\it (b) Dipartimento di Fisica, Universit\`a degli Studi
di Perugia, via A. Pascoli
06100 Perugia, Italy, and INFN, sezione di Perugia }
\end{center}

\vskip 1.0cm
\centerline{\bf Abstract}
\vskip 0.4cm
According to recent studies,
Parton Distributions Functions (PDFs) and
Generalized Parton Distributions (GPDs)
can be evaluated in a Constituent Quark Model (CQM) scenario,
considering the constituent quarks as composite objects.
In here, a fully covariant model for a system
of two particles, together with its non relativistic limit,
are used to calculate PDFs and GPDs.
The analysis permits to realize that by no means
the effects of Relativity can be simulated
taking into account the structure of the
constituent particles, the two effects being
independent and necessary for a proper description
of available high energy data in terms of CQM.

\vskip 1cm
\leftline{Pacs: 12.39-x, 13.60.Hb, 13.88+e}
\leftline{Keywords:  hadrons, partons, generalized parton distributions,
quark models.}
\vspace{.7cm}

{\tt
\leftline{santiago.noguera@uv.es}
\leftline {sergio.scopetta@pg.infn.it}
\leftline{vicente.vento@uv.es}
}


\end{titlepage}

\section{Introduction}

Parton Distributions Functions (PDFs) and
Generalized Parton Distributions (GPDs)
\cite{first,radnew,ji1},
the latter relating PDFs and electromagnetic
Form Factors (FF), encode
unique information on the non perturbative
hadron structure (for a recent review, see \cite{dpr}).
In principle,
any realistic model of hadron structure
should be able to estimate them.
Among the non perturbative approaches,
the Constituent Quark Model
(CQM) has a long story of successful
predictions in low energy studies of the electromagnetic
structure of the nucleon, such as the calculation of FF.
In the high energy sector,
in order to compare model predictions of PDFs
with Deep Inelastic Scattering (DIS) data,
one has to evolve, according to perturbative QCD, the leading twist
component of the physical structure functions obtained
at the low momentum scale associated with the model.
Such a procedure, already addressed in \cite{pape,jaro},
has proven successful in describing the gross features of
standard PDFs by using different CQM (see, e.g., \cite{trvv}),
and it has been applied also to the calculation of
the valence quark contribution to GPDs in Ref. \cite{epj}.
Anyway, in order to achieve a better agreement with data, such
a program has to be improved.

Two main directions have been followed in this respect by
different authors. One has been to show that unpolarized and
polarized DIS data are consistent with a low energy scenario
dominated by composite constituent quarks of the nucleon
\cite{scopetta1}. The latter are defined through a scheme
suggested by Altarelli, Cabibbo, Maiani and Petronzio (ACMP)
\cite{acmp}, updated with modern phenomenological information. The
idea of complex constituents, as old as the quark-parton model
itself \cite{morp}, used extensively in other frameworks
\cite{hwa}, has been recently applied to demonstrate the evidence
of complex objects inside the proton, analyzing intermediate
energy data of electron scattering off the proton \cite{psr}. In
Ref.\cite{prd}, the idea has been successfully applied also to
GPDs, in particular allowing for the evaluation of the sea quark
contribution, so that GPDs can be calculated in their full range
of definition. Such an achievement will permit to estimate the
cross-sections which are relevant for actual GPDs measurements,
providing us with an important tool for planning future
experiments. In any case, one has to realize that the CQM
calculations, as developed in \cite{trvv}, even if the structure
of the constituent is taken into account as in \cite{scopetta1},
are affected by the problem of poor support, being the Bjorken
variable $x_{Bj}$ not limited between zero and one.

The other direction makes use of light-front dynamics, which
allows to estimate relativistic effects in a covariant framework.
Another good feature of this program is that, by construction, it
is not affected by the problem of poor support. This approach is
particularly useful when spin degrees of freedom are considered,
and it has been indeed applied to the calculation of polarized,
transversity and orbital angular momentum distributions
\cite{pietro}. Recently, it has been also applied to the
calculation of the quark contribution to spin independent and spin
dependent GPDs \cite{bpt}, using the overlap representation
\cite{kroll}. A relevant contribution to the calculation of GPDs
on the light-front has been given by Tiburzi and Miller
\cite{mill}, and some remarks on the use of light-front for
calculating GPDs can be found in \cite{sil}

A question which naturally arises is whether or not the two
approaches described above, the one which takes into account the
structure of the constituent quark, or the one which implements
relativity in a CQM by a light front approach, are introducing in
different effective ways the same physics into the problem. In
other words, whether or not the structure of the constituent quark
has to be implemented even in a relativistic model, such as the
one obtained in a light front approach. These issues are discussed
in the present paper. To do so we implement the discussion in a
simple model of a bound system of two scalar particles, defined in
a quantum field theoretical, explicitly covariant, framework
\cite{lukas}. The model, despite of its simplicity, is a rather
general one and allows the exact evaluation of the PDF's and
GPD's. Moreover, their Non Relativistic (NR) limit can be
calculated. We will show that it is not possible to recover the
shape of the initial full covariant PDFs and GPDs by implementing,
in the distributions obtained as their NR limit, the structure of
the constituent quark. We find therefore that, for the quite
general model under scrutiny, the effect of introducing the
structure of the constituents describes different physics than
that described by implementing Relativity.

The paper is structured as follows.
After a short definition of the main quantities
of interest, the used model and its NR limit
are presented in section II, together with
the possible modifications due to the structure
of the constituent quark;
in section III, the results of the calculations are shown.
Conclusions are drawn in the last section.

\section{Formalism}

PDFs and GPDs  are the main quantities of interest
in this paper. Here below they are shortly defined
together with the conventions used.
The GPDs are non-diagonal matrix elements of bi-local field
operators.
Let us consider a scalar system of mass $M$, with
initial momentum $P$,
final momentum $P^{\prime }$, and momentum transfer given by
$\Delta =P^{\prime }-P$, made of two scalar
particles of mass $m$. If the momentum of the interacting one is labeled
by $p$ and the quantity $\bar P = (P+P')/2$ is used,
the GPD of such a system is defined
by the matrix elements of bi-local scalar field
operators~\cite{first,radnew,ji1}:
\begin{equation}
{\mathcal{J}}^{+}\equiv \frac{1}{2}\left. \int \frac{dz^{-}}{2\pi }%
e^{ixP^{+}z^{-}}\left\langle P^{\prime }\right\vert \Phi ^{\dagger }\left(
0\right)
{\lrpartial}{}^{+}\Phi \left( z\right)
\left\vert P\right\rangle \right\vert _{z^{+}=z^{\perp }=0}={{\mathcal{H}}}%
(x,\xi ,t),  \label{spdgen}
\end{equation}
In the above equation, $\xi $ is the so-called
skewedness parameter, defined as
\begin{equation}  \label{ratios}
\xi=-\frac{\Delta^+}
{2 \bar P^+}~,
\end{equation}
so that
\begin{equation}  \label{ratio1}
x + \xi =\frac{p^+}{\bar P^+}~,
\end{equation}
and $
{\lrpartial }=
{\rpartial }-
{\lpartial }$
(any four vector $v^\mu$ will be denoted $%
(v^+,v^{\perp},v^-)$, where the light cone variables are defined by $v^\pm =
(v^0\pm v^3)/\sqrt{2}$ and the transverse part $v^{\perp}=(v^1,v^2)$).
The standard PDF is defined as the forward limit of
Eq. (\ref{spdgen}), when $t \rightarrow 0$ an $\xi \rightarrow 0$.

The elastic
electromagnetic form factor of a system composed of two scalar particles is
given by:
\begin{equation}
J^{+}\equiv \left\langle P^{\prime }\right\vert \Phi ^{\dagger }\left(
0\right)
{\lrpartial}{}^{+}\Phi \left( 0\right)
\left\vert P\right\rangle =(P+P^{\prime })^{+}\,F(t).  \label{emffgen}
\end{equation}
It follows directly from these definitions that integrating the GPD over $x$
gives the form factor,
\begin{equation}
\int {{\mathcal{H}}}(x,\xi ,t)\,dx=F(t),  \label{relat}
\end{equation}
where the dependence on the skewedness parameter $\xi $ drops
out. This result is an important constraint for any model
calculation.

Only elastic processes will be considered, so $P^{2}=P^{\prime
2}=M^{2} $ and $\Delta ^{2}=t$.
The values of $\xi$ which are possible for a given value of
$\Delta^2$ are:
\bq
0 \le \xi \le \sqrt{- \Delta^2}/\sqrt{4 M^2-\Delta^2}~.
\label{xim}
\eq

In Ref. \cite{lukas}, GPDs have been estimated
by a simple model, which allows for a
completely analytic solution of the Bethe-Salpeter equation.
The model
describes a bound state of two distinguishable equal-mass scalar particles
bound together by a zero-range interaction.
The Lagrangian is,
\begin{equation}
{\mathcal{L}}=\left[ D_{\mu }\phi \right] ^{\dagger }\left[ D^{\mu }\phi %
\right] -m^{2}\phi ^{\dagger }\phi +\frac{1}{2}\partial _{\mu }\chi \partial
^{\mu }\chi -\frac{1}{2}m^{2}\chi ^{2}-\frac{g}{2}\left( \phi ^{\dagger
}\phi \,\chi ^{2}\right) ,  \label{ZRlag}
\end{equation}
with $D_{\mu }=\partial _{\mu }+ieA_{\mu }$ so that the electromagnetic
charge only couples to the field $\phi $.
Being
the coupling
constant $g$ larger than a critical value, bound states are encountered.
The corresponding Bethe-Salpeter
equation can be trivially solved in the ladder
approximation.
The theory
is renormalizable and a renormalization program for bound states
can be defined.

The main advantage of the model defined by Eq.~(\ref{ZRlag})
lies in its simplicity, in the fact that
one may obtain analytic solutions, avoiding approximations that might
destroy physical requirements, symmetries
or sum rules.
These properties make
it a useful playground to perform benchmark calculations, as it was used
recently in order to test the viability of certain relativistic quantum
mechanics approaches~\cite{desp}.

In this model, the GPD ${\mathcal{H}}$ is obtained as an integral
over Bethe-Salpeter amplitudes. It reads:
\begin{equation}
{{\mathcal{H}}}(x,\xi ,t)=
x\,\frac{C^{2}}{i}\int \frac{d^{4}p%
}{(2\pi )^{4}}\frac{\delta (x + \xi
-p^{+}/\bar P^{+})}{(p^{2}-m^{2}+i\epsilon )\left[
(p+\Delta )^{2}-m^{2}+i\epsilon \right] \left[ (P-p)^{2}-m^{2}+i\epsilon %
\right] }.  \label{spdfbsp}
\end{equation}
Inspecting the pole structure of the integrand for the evaluation of the $%
p^{-}$ integral, one realizes that it
vanishes unless $- \xi\leq x\leq 1$, i.e., the
GPDs have the correct support properties. The integral of Eq.~(\ref{spdfbsp})
may be calculated analytically and the explicit result
is given in Ref. \cite{lukas}, where it is shown
that it fulfills the polynomiality condition \cite{jig}.
The quark distribution function is very simple and
is written again here below for the reader's convenience:
\begin{equation}
q(x)\equiv {{\mathcal{H}}}(x,0,0)=\frac{C^{2}}{16\pi ^{2}}\frac{x(1-x)}{%
m^{2}-x(1-x)M^{2}}.  \label{zrqdf}
\end{equation}
The normalization integral may be done analytically and determines the
normalization constant $C$; it is clear that crucial parameters
of the model are the mass $M$ of the hadron and the
mass $m$ of its constituents.

Among the good properties of the model,
use will be done here of the
fact that it allows for a clear NR limit.
This is done by considering a NR approximation of the energies
appearing in the denominator of Eq. (\ref{spdfbsp}), i.e.
by taking, up to order $O( \vec p^2/m^2 )$ :
\begin{eqnarray}
p_0 & \simeq &  m + {\vec p^2 \over 2 m}~,
\nonumber
\\
(P-p)_0 & \simeq &  m + {(\vec P - \vec p)^2 \over 2 m}~,
\nonumber
\\
(p+\Delta)_0 & \simeq &  m + {(\vec p + \vec \Delta)^2 \over 2 m}~.
\end{eqnarray}
One should realize that the above approximations
also imply, as already discussed in \cite{prd},
$-t \ll m^2$, $\xi^2 \ll 1$.
This means that the NR limit in the calculation of GPDs
is rougher than in the calculation of standard PDFs,
because it implies an additional approximation
on the momentum transfer.

Neglecting some terms which are
of order $O( \vec p^2 / m^2 )$, one finds:

\begin{eqnarray}
{{\mathcal{H}}}(x,\xi,t) & = &
{ C^2 \over (2 \pi)^3} \, x \, \bar M \int _{p_{min}(x,t)}^{p_{max}(x,t)}
dp \, p
\nonumber
\\
& \times &
\int_0^{2 \pi}
{  d \phi
 \over
2 \left ( m + { p^2 \over 2m} \right )
\{ [ \bar M ( \bar M - 2 ( m + { p^2 \over 2 m } ))
+ {t \over 4} ]^2 -d^2        \} }~,
\label{nr}
\end{eqnarray}

where $\bar{M} = \sqrt{1 - t/4}$,
$d = \vec \Delta_\perp \cdot \vec p_\perp - 2 \bar M \xi \tau p $,
$\tau = [\bar M (1 - x) - m - p^2/(2m)]/p$,
$p_{max}(x,t)= m (1+A)$, $p_{min}(x,t)=Max \{ m(-1 + A), m(1 - A)  \}$,
$A= \sqrt{2(\bar M / m) (1-x) - 1 }$.

The forward limit is again analytical, being given by:
\begin{equation}
q(x)\equiv {{\mathcal{H}}}(x,0,0)=
{ C^2 \over 2 (2 \pi)^2}\, {m^3 \over M} \, x \,
\left \{
I(\,p_{max}^2(x,0)\,) -
I(\,p_{min}^2(x,0)\,)
\right \}~,
\label{nrf}
\end{equation}
where
\begin{equation}
I(y) = - {1 \over (y-a) (a-b)} -
{ 1 \over (a-b)^2} \ln \left | {y-a \over y-b } \right |~,
\end{equation}
with $a = m (M-2m)$ and $b = -2m^2$.

In Ref. \cite{prd}, a convolution formula
has been derived, giving
the quantity ${\mathcal{H}}_q$, i.e. the contribution of the
quark of flavor $q$ to the GPD ${\mathcal{H}}_q$,  in terms of a
constituent quark off-forward momentum
distribution, ${\mathcal{H}}_{q_0}$,
and of a GPD of the constituent quark $q_0$ itself,
${\mathcal{H}}_{q_0q}$.
It is assumed that the hard scattering with the
virtual photon takes place on a parton of a hadron target, made of
complex constituents, in an Impulse Approximation scenario.
One parton ({\sl{current}}) quark,
belonging to a given constituent, interacts with the probe
and it is afterwords reabsorbed by the same constituent,
without further re-scattering with the recoiling system.
Details of the approach can be found in Ref. \cite{prd,prc}.

The convolution formula, valid for low values of $t$ and $\xi$,
can be written in the form \cite{prd}:
\begin{eqnarray}
{\mathcal{H}}_{q}(x,\xi,t) =
\sum_{q_0} \int_x^1 { dz \over z}
{\mathcal{H}}_{q_0}(z, \xi ,t )
{\mathcal{H}}_{q_0 q} \left( {x \over z},
{\xi \over z},t \right )~,
\label{main}
\end{eqnarray}
where ${\mathcal{H}}_{q_0}$ is the GPD to be evaluated in any
CQM, such as the scalar model under scrutiny here,
for the flavor $q_0$, while ${\mathcal{H}}_{q_0 q}( {x \over z},
{\xi \over z},t)$ is the constituent quark
GPD.
One can realize that the GPD defined by Eq.
(\ref{main}) satisfies the polynomiality condition
if both the functions ${\mathcal{H}}$
and ${\mathcal{H}}_{q_0 q}$ do it.
This will be the case for the distributions used
in this paper.

The constituent quark GPD
${\mathcal{H}}_{q_0 q}( {x \over z},{\xi \over z},t)$
has also been modelled in \cite{prd}.
As usual, one can start modelling this quantity
thinking first of all to its forward limit,
where the constituent quark
parton distributions have to be recovered.
As already said in the previous section,
in a series of papers
a simple picture of the constituent quark as a
complex system of point-like partons
has been proposed
\cite{scopetta1},
re-taking a scenario suggested by Altarelli,
Cabibbo, Maiani and Petronzio (ACMP)
\cite{acmp}.

According to that idea,
the structure of the constituent quark
is described by a set of functions
$\phi_{q_0q}(x)$ that specify the number of point-like partons
of type $q$ which
are present in the constituent of type $q_0$, with fraction $x$ of its total
momentum. These functions will be called, generically, the structure
functions of the constituent quark. They are expressed in
terms of the independent $\phi_{q_0q}(x)$ and of the constituent
density distributions ($q_0=u_0,d_0$) as,
\bq
q(x)=\sum_{q_0}\int_x^1 {dz\over z}
q_0(z) \phi_{q_0q} \left({x \over z} \right)~,
\label{fx}
\eq
where $q$ labels the various partons, i.e., valence quarks
($ u_v,d_v$), sea quarks ($u_s,d_s, s$), sea antiquarks ($\bar u,\bar d, \
\bar
s$) and gluons $g$.
The different types and functional forms of the structure functions of the
constituent quarks are derived from three very natural assumptions, i.e.
the point-like partons are $QCD$ degrees of freedom,
i.e. quarks,
antiquarks and gluons; Regge behavior for $x\rightarrow 0$ has
to be valid; invariance under charge conjugation and isospin
has to be reproduced.
The last assumption of the approach relates to the scale at
which the constituent quark
structure is defined. We choose for it the so called hadronic
scale $\mu_0^2$ \cite{trvv,grv}. This hypothesis fixes $all$ but one
the parameters of
the approach. The only free one is fixed according to
the value of $F_2$ at $x=0.01$ \cite{acmp},
and its value is chosen again according to \cite{grv}.
We stress that all these inputs are forced only by the updated
phenomenology, through the 2$^{nd}$ moments of PDFs.
The values of the parameters obtained are listed in \cite{scopetta1}.

These considerations define, in the case of the valence quarks, the following
structure function

\bq
\phi_{q_0 q_v}({x})
= { \Gamma(A + {1 \over 2}) \over
\Gamma({1 \over 2}) \Gamma(A) }
{ (1-x)^{A-1} \over \sqrt{x} }.
\label{csf1}
\eq

The physical arguments leading to this expression are
indeed the ones listed above. In fact,
first of all, Regge theory fixes the
behavior $x^{-1/2}$ for $x \rightarrow 0$. Second, the
function has to be normalized to 1, because any constituent
has to contain a leading valence current quark
with the same quantum numbers.
This latter fact fixes the constant in front
of the function in the above equation. Eventually, the
parameter $A$ is fixed to the value 0.435, by imposing that
the second moment at the low scale of the model is
reproduced.
The corresponding structure functions for the sea
and gluons, not used here, can be found in
\cite{scopetta1,prd}.

One should realize that the original model was thought
for spin $1/2$ constituent quarks, while here scalar constituents
are discussed. Anyway, since only spin-independent
observables will be evaluated, there is no physical reason
to change the structure functions of the model.

This scenario has been generalized in Ref. \cite{prd}
to describe off-forward phenomena.
The main steps are reported here.
First of all, the forward limit of the
GPDs formula, Eq. (\ref{main}), has to be given by
Eq. ({\ref{fx}}).
By taking the forward limit of Eq. (\ref{main}), one obtains:
\begin{eqnarray}
{\mathcal{H}}_{q}(x,0,0) & = &
\sum_{q_0} \int_x^1 { dz \over z}
{\mathcal{H}}_{q_0}(z, 0, 0 )
{\mathcal{H}}_{q_0 q} \left( {x \over z},0,0 \right)
\nonumber \\
& = &
\sum_{q_0} \int_x^1 { dz \over z}
q_0(z)
{\mathcal{H}}_{q_0 q} \left( {x \over z},0,0 \right)~,
\label{for}
\end{eqnarray}
so that, in order for the latter to coincide with Eq.
(\ref{fx}), one must have
${\mathcal{H}}_{q_0 q}( x,0,0) \equiv
\phi_{q_0q}(x)$.
In such a way, through the ACMP prescription, the forward
limit of the unknown constituent quark GPD
${\mathcal{H}}_{q_0 q}( {x \over z},
{\xi \over z},t)$ can be fixed.
The off-forward behavior of the Constituent Quark GPDs
can be modelled in a natural way by using the ``$\alpha$-Double
Distributions'' (DD's) language proposed by Radyushkin \cite{radd}.
DD's, $\Phi(\tilde x, \alpha, t)$,
are a representation of GPDs which automatically guarantees
the polynomiality property.

The relation between any GPD ${\mathcal{H}}$, defined {\sl \`a la Ji},
for example the one we need, i.e.
${\mathcal{H}}_{q_0q}$ for the constituent quark target,
is related to the $\alpha$-DD's, which we call
$\tilde \Phi_{q_0 q} (\tilde x, \alpha,t)$ for the constituent
quark,
in the following way
\cite{radd}:
\begin{eqnarray}
{\mathcal{H}}_{q_0q}(x,\xi,t) = \int_{-1}^1 d\tilde x
\int_{-1 + |\tilde x|}^{1-|\tilde x|}
\delta(\tilde x + \xi  \alpha - x)
\tilde \Phi_{q_0 q} (\tilde x, \alpha,t) d \alpha~.
\label{hdd}
\end{eqnarray}

With some care, the expression above can be integrated over
$\tilde x$ and the result is explicitly given in
\cite{radd}.
The DDs fulfill
the polynomiality condition \cite{jig}.

In \cite{radd}, a factorized
ansatz is suggested for the DD's:
\begin{equation}
\tilde \Phi_{q_0 q} (\tilde x, \alpha,t) =
h_{q} (\tilde x, \alpha,t)
\Phi_{q_0 q} (\tilde x)
F_{q_0}(t)~,
\label{ans}
\end{equation}
with
the $\alpha$ dependent
term,
$h_{q} (\tilde x, \alpha,t)$,
which has the character of a mesonic amplitude.
Besides, in Eq. (\ref{ans})
$
\Phi_{q_0 q} (\tilde x)
$
represents the forward density
and, eventually,
$
F_{q_0}(t)
$
the constituent quark form factor.
It can be easily verified that the GPD of the constituent quark,
Eq. (\ref{hdd}), with the factorized form Eq. (\ref{ans}),
fulfills the
crucial constraints of GPDs, i.e., the forward limit,
the first-moment and the polynomiality condition,
the latter being automatically verified in the DD's
description.
In the following the
above factorized form will be assumed,
so that we need to model the three functions
appearing in Eq. (\ref{ans}).

For the amplitude $h_q$, use will be made of
one of the simple normalized forms suggested in
\cite{radd},
on the bases of the symmetry
properties of DD's (see \cite{prd}).

Besides, since we will identify
quarks for $x \ge \xi/2$, pairs for $ x \le |\xi/2|$,
antiquarks for  $x \le -\xi/2$, and, since
in our approach the forward densities
$\Phi_{q_0q} (\tilde x)$ have to be given by
the standard $\Phi$ functions of the
$ACMP$ approach,
one has, for the DD of flavor $q$
of the constituent quark:
\begin{eqnarray}\label{sr}
\tilde \Phi_{q_0 q} (\tilde x, \alpha,t) =
\cases{
(h_{q}(\tilde x,\alpha) \Phi_{q_0q_v}(\tilde x)
+
h_{q}(\tilde x,\alpha) \Phi_{q_0q_s}(\tilde x)) F_{q_0}(t)
 &
for $\tilde x \ge 0$ \cr
- h_{q}(- \tilde x,\alpha) \Phi_{q_0q_s}(- \tilde x)
F_{q_0}(t)
 & for $\tilde x < 0$ \cr}
\label{dd}
\end{eqnarray}

Eventually, as a f.f. we will take a monopole form
corresponding to a constituent quark size $r_Q \simeq 0.3 fm$:

\begin{equation}
F_{q_0}(t) = { 1 \over 1 - {1
\over 6 }
r_Q^2 t
}~,
\label{ffacmp}
\end{equation}
a scenario strongly supported by the analysis of \cite{psr}.

By using such a f.f., the amplitude $h_q$ and the standard
ACMP $\Phi$'s,
in Eq. (\ref{dd}),
and inserting the obtained
$\Phi_{q_0 q} (\tilde x, \alpha,t)$
into Eq. (\ref{hdd}),
the constituent quark GPD in the ACMP scenario can be eventually calculated.

All the ingredients of the calculation have therefore been
introduced. In the next section,
results will be shown for the GPD ${\mathcal{H}}$,
calculated in the scalar model according to Eq. (\ref{spdfbsp}) ,
for its NR limit, evaluated by means of Eq. (\ref{nr}),
and for their forward limit, also considering the structure of the
constituent described above.

\section{Results and discussion}

In this sections we show the results of our calculation in a
series of figures and discuss their implication in physical terms.

In Fig. 1, the PDF obtained with the scalar model, Eq.
(\ref{zrqdf}), is shown together with its NR limit, Eq.
(\ref{nrf}). The mass of the constituent is taken to be $m=$ 0.240
MeV, while the mass of the bound system is fixed to $M=0.432$ MeV.
This corresponds to a binding energy of 48 MeV, i.e. 20 \% of the
constituent mass. The system defined in this way is therefore
quite relativistic. It is seen that the NR limit does not
reproduce the high momentum tail of the exact distribution.
Moreover, it is has poor support. This defect can be quantified by
measuring the second moment, once the first has been fixed to 1.
The second moment of the exact PDF gives 0.5, as it
should. Instead, the NR distribution gives 0.444, so that a
violation of the order of 10\%, due to the poor support, is found.

Fig. 2 shows the same results but for a hadron mass of $M=0.475$
MeV, i.e. a system with a binding energy 2 \% the mass of the
constituent, and weakly bound, essentially a NR one. One should
recall that a system like this is still more bound than any atomic
nucleus, by a factor of two, approximately. In this case the
amount of support violation is only of the order of 1\% since the
the second moment sum rule gives 0.488. This observation supports
the use of the Impulse Approximation and of NR wave functions for
the estimates of the nuclear parton distributions. In any case, it
is evident that also in this case the NR approximation is not able
to reproduce the high momentum tail of the quark distribution, the
problem being anyway less serious than for the relativistic
situation shown in Fig. 1.

From Figs 1 and 2 it is evident that, in order to describe
more relativistic distributions by means of the scalar model
under scrutiny, it is enough to increase the binding of the system,
i.e., the mass of the system has to be reduced keeping fixed the
mass of the constituents.

In Fig. 3, we show the effect of considering the structure of the
constituents, in the deeply bound scenario of Fig. 1. Eq.
(\ref{main}) has been evaluated in the forward limit, by using Eq.
(\ref{csf1}) for the structure function of the constituents,
${\mathcal{H}}_{q_0q}$, and using Eq. (\ref{zrqdf}) and Eq.
(\ref{nrf}) as ${\mathcal{H}}_{q_0}(x,0,0)$. The two curves are
shown together with the result obtained without considering the
structure of the constituents, given simply by Eqs. (\ref{zrqdf})
and (\ref{nrf}). It is seen that the effect of inserting some
structure for the constituents in the NR model produces does not
help to reproduce the high momentum components dropped in the
while performing the NR limit. Quite on the contrary, the ACMP
structure increases the number of low-x current quarks. One should
also notice that, even changing the parameters of the constituent
structure functions, it is not possible to simulate the
relativistic result, unless the physical arguments used to build
Eq. (\ref{csf1}) are obviated. We stress that these arguments are
quite general based on pQCD and Regge theory arguments. To drop
them would be equivalent to consider questionable Regge theory or
the capability of pQCD to predict the evolution of the second 
moment of the valence quark distribution. All the curves shown in Fig.3 are
multiplied by $x$, for the sake of clarity.

Fig. 4 is just an illustration of the full procedure required to
describe DIS data starting from CQM. First of all, the parton
distribution is to be calculated in a (relativistic) model.
Relativity is necessary especially if one wants to evaluate GPDs
at large $t$ and $\xi$. After that, some structure for the
constituents, which fixes the scale of the model calculations
(here it turns out to be $\mu_0^2 = 0.34$ GeV$^2$), should be
considered. Finally,  pQCD evolution of the model result up to the
experimental scale should be performed. At this point, the model
predictions can be compared with data. In Fig. 4, the valence
quark distribution of the pion, extracted from data at $Q^2$= 4
GeV$^2$ \cite{mrst}, is compared with the result of the scalar
model calculation, once the structure is taken into account and
the evolution is performed. The mass of the constituent is taken
to be $m=0.240$ MeV, as always, while the mass of the hadron is
chosen to be $M=0.140$ MeV,
close to the physical pion mass.
Considering that the scalar model is basically used here as toy
model, the exercise produces an unexpected good agreement with
data. We do not claim that a system like the pion can be described
by a model like the one under scrutiny. However, from the
agreement found draw some conclusions: i) the model used, despite
its simplicity, has many good features which makes its use to
study hadron structure physically meaningful; ii) the structure of
the constituent, given by Eq. (\ref{csf1}), kept unchanged
throughout our investigations \cite{scopetta1,prd}, seems quite
general and useful in varied situations; iii) if one starts from a
CQM, relativity and structure of the constituents have to be
considered simultaneously to be able to describe the data. To
emphasize the latter point,  recall that relativity strongly
changes the large $x$ region, while the structure of the
constituents mainly affects the small $x$ region. The issue of a
detailed study of the pion DIS structure function, starting from a
more realistic CQM, as has been done by several groups
\cite{arr,acmp1}, is beyond the scope of the present paper and
will be discussed elsewhere.

Results in the non forward case, at low values
of $t$ and $\xi$, are shown in Figs. 5 and 6.

In Fig. 5, the GPD in the scalar model, obtained from Eq.
(\ref{spdfbsp}), is compared with its NR limit, Eq. (\ref{nr})
again in the deeply bound scenario of Fig. 1. It is seen that the
poor support becomes more serious with respect to the forward
case.

In Fig. 6, the same is shown once the structure has been taken
into account, according to Eq. (\ref{main}). The same conclusions
as for the forward case, i.e. the different nature of the effects
due to relativity and constituent quark structure, can be drawn.

\section{Conclusions}

In this paper, a fully covariant model for a scalar system of two
scalar particles is used as a physically meaningful toy model to
calculate Parton Distributions Functions and Generalized Parton
Distributions. The analysis permits to check the conclusions of
recent studies, according to which parton distributions can be
evaluated in a Constituent Quark Model scenario,  considering the
constituent quarks as composite objects, developing an idea which
dates back to the seventies. The NR limits of the corresponding
distributions are also evaluated. The analysis shows that the
effects of Relativity cannot be simulated by the structure
proposed for the constituent particles, which is based on quite
general physical arguments. The two effects are found to be
independent and both necessary for a proper description of
available high energy data in terms of CQM.

\section{Acknowledgments}

S.S. thanks the Department of Theoretical Physics of the Valencia
University, where this work has been done, for a warm hospitality
and financial support. This work is supported in part by
MCyT-FIS2004-05616-C02-01, GV-GRUPOS03/094~ and by MIUR through
the funds COFIN03.

\newpage

\begin{figure}[ht]
\vspace{6.6cm}
\includegraphics{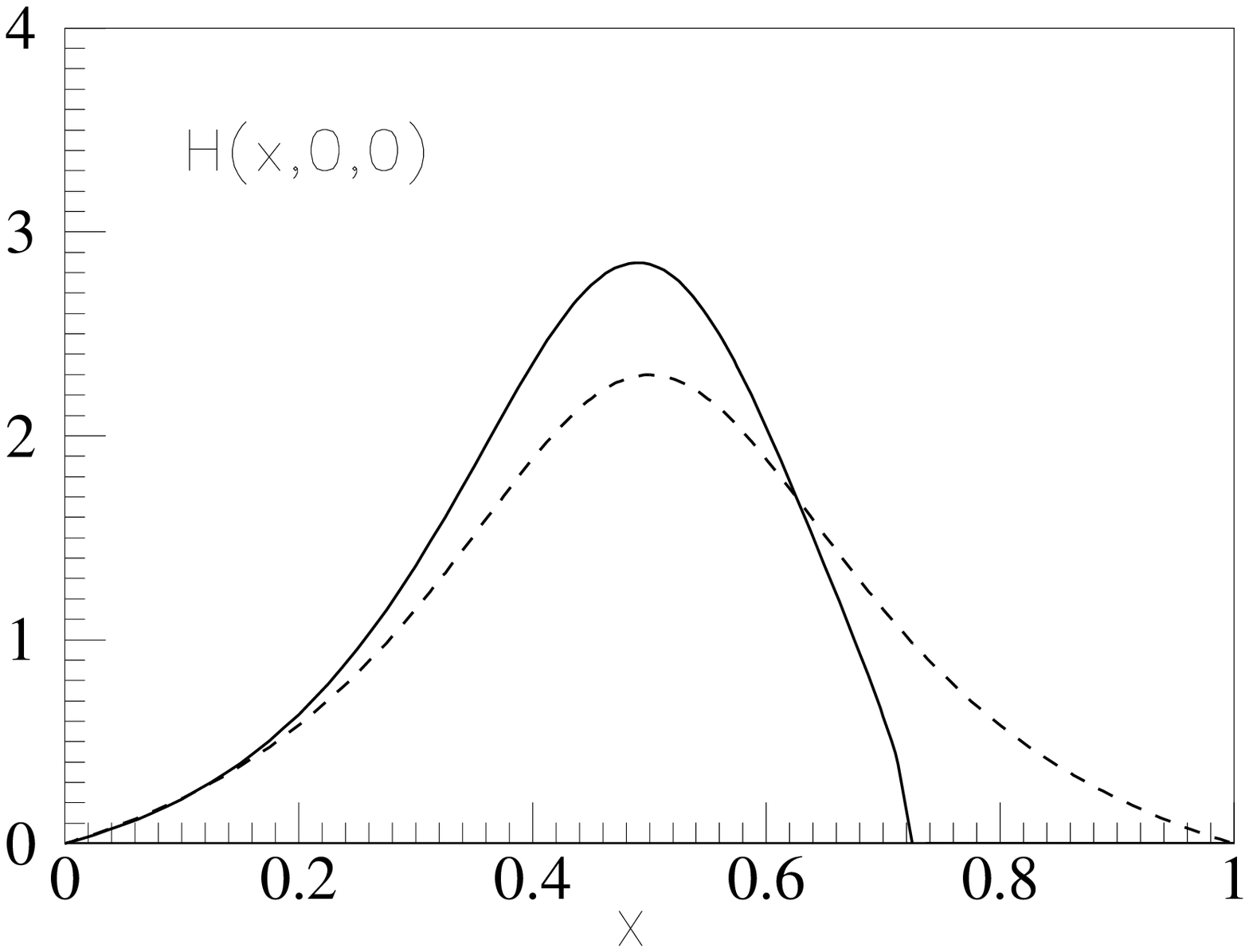}
\caption{
The PDF obtained with the scalar model,
Eq. (\ref{zrqdf}) (dashed), is shown together
with its NR limit, Eq. (\ref{nr}) (full).
The mass of the constituent is taken to be
$m=$ 0.240 MeV, while the mass of the bound system
is fixed to $M=$ 0.432 MeV. }
\end{figure}

\begin{figure}[ht]
\vspace{6.6cm}
\includegraphics{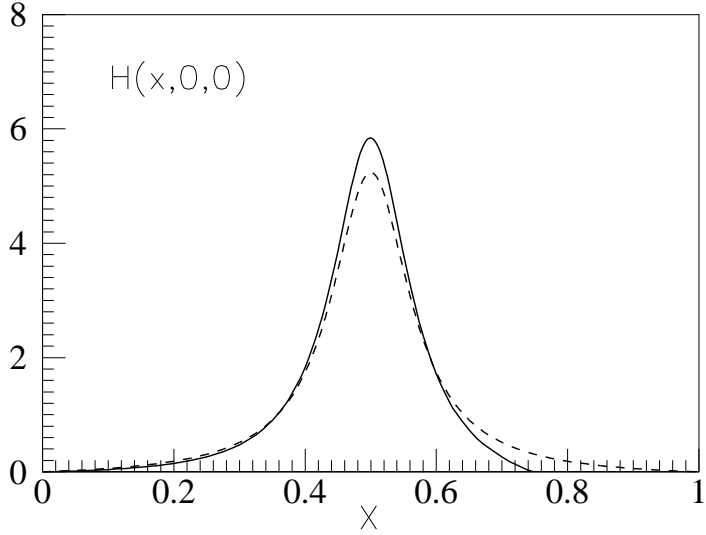}
\caption{
The same as in Fig. 1 but for constituents of mass
$m=0.240$ MeV and a bound system of mass $M=0.475$ MeV.
}
\end{figure}

\newpage

\begin{figure}[ht]
\vspace{6.8cm}
\includegraphics{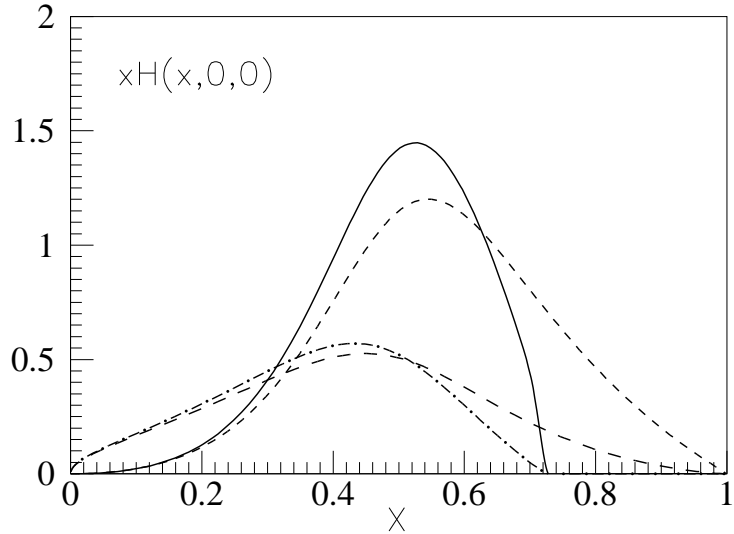}
\caption{
The long-dashed curve represents
Eq. (\ref{main}), evaluated in the forward limit,
using Eq. (\ref{csf1}) for the structure function of the constituent,
${\mathcal{H}}_{q_0q}$, and Eq. (\ref{zrqdf}) for ${\mathcal{H}}_{q_0}(x,0,0)$.
The dot-dashed curve represents
Eq. (\ref{main}) evaluated in the forward limit,
using Eq. (\ref{csf1}) for the structure function of the constituent,
${\mathcal{H}}_{q_0q}$, and Eq. (\ref{nrf}) for ${\mathcal{H}}_{q_0}(x,0,0)$.
Dashed curve is given by Eq. (\ref{zrqdf}),
the full curve by Eq. (\ref{nrf}).
All the Equations have been multiplied by $x$
to give the curves which are shown.
The mass of the constituent is
$m=0.240$ MeV, while the mass of the bound system
is $M=0.432$ MeV.
}
\end{figure}

\vskip 2cm

\begin{figure}[ht]
\vspace{6.6cm}
\includegraphics{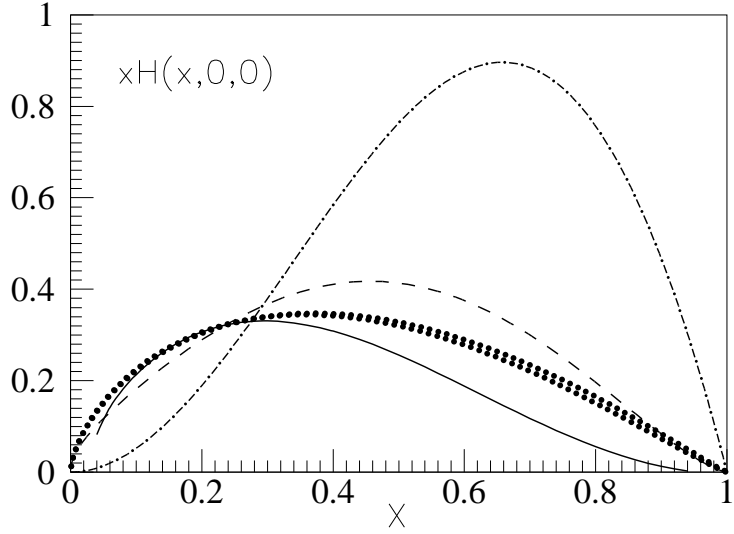}
\caption{
The dot-dashed line is the result of the scalar model, in the forward limit,
Eq. (\ref{zrqdf}), for $M=0.140$ MeV and $m=0.240$ MeV; the dashed line
is obtained by considering the structure of the constituent, i.e.
inserting the Eq. (\ref{csf1}), together
with Eq. (\ref{zrqdf}), into Eq. (\ref{main}).
The full line represents the evolution, up to $Q^2 = 4$ GeV$^2$,
of the dashed curve, the latter assumed to be valid
at a scale of $\mu_0^2 = 0.34$ GeV$^2$.
All the Equations have been multiplied by $x$, to
give the curves which are shown.
The dots represent the data for the valence quark distribution
in the pion at $Q^2 = 4$ GeV$^2$, multiplied by
$x$, as it is parameterized in
\cite{mrst}, with their uncertainties.
}
\end{figure}

\newpage

\begin{figure}[ht]
\vspace{6.8cm}
\includegraphics{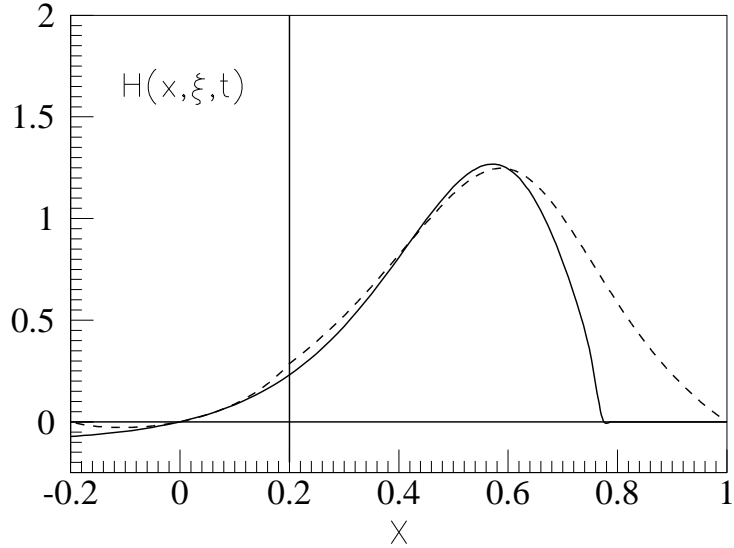}
\caption{
The GPD in the scalar model, obtained from
Eq. (\ref{spdfbsp}) at $t=-0.3$ GeV$^2$ and
$\xi = 0.2$ (dashed), compared with its NR limit,
Eq. (\ref{nr}) (full).
The mass of the constituent is
$m=0.240$ MeV, while the mass of the bound system
is $M=0.432$ MeV.
}
\end{figure}

\vskip 2cm

\begin{figure}[ht]
\vspace{6.6cm}
\includegraphics{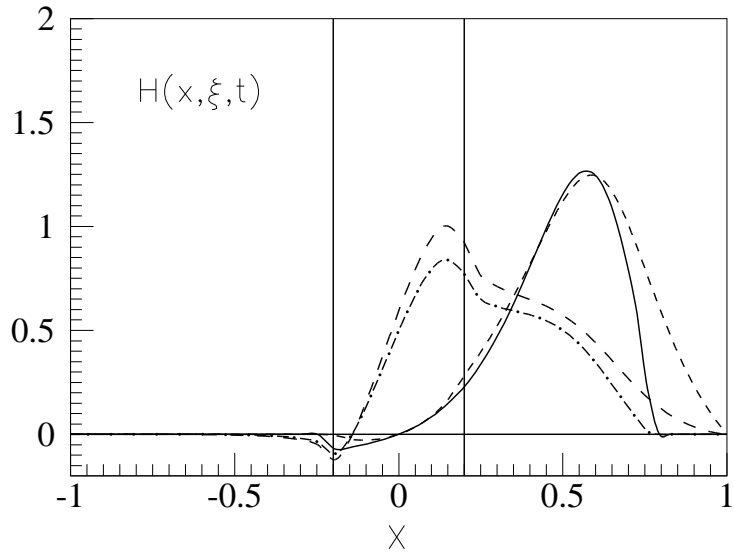}
\caption{
The GPD in the scalar model, obtained from
Eq. (\ref{spdfbsp}) at $t=-0.3$ GeV$^2$ and
$\xi = 0.2$ (dashed), compared with its NR limit,
Eq. (\ref{nr}) (full).
The long-dashed and dot-dashed lines
are obtained from the two previous ones,
respectively, taking into account the structure
of the constituent quark, according to Eq. (\ref{main}).
The mass of the constituent is
$m=0.240$ MeV, while the mass of the bound system
is $M=0.432$ MeV.
}
\end{figure}

\end{document}